\documentclass[reprint,amsmath,amssymb,aps,prd]{revtex4-2}
\pdfoutput=1
\usepackage{filecontents}
\usepackage{graphicx}
\usepackage{dcolumn}
\usepackage{bm}
\usepackage[usenames,dvipsnames]{color}
\usepackage{longtable}
\usepackage{tensor}
\usepackage{epsfig}
\usepackage{amssymb}
\usepackage{amsmath}
\usepackage{verbatim}
\usepackage{xspace}
\usepackage{aas_macros}
\include{jdefs}
\usepackage{graphics,hyperref}
\newcommand{\wt}{{\hspace{-0.05cm}\scalebox{0.65}{\textit{T}}}}

\begin{document}
\title{Non-symmetric wormholes and localized Big Rip singularities in Einstein-Weyl gravity}
\author{A.~Bonanno${}^{1,2}$, S.~Silveravalle${}^{3,4}$ and A.~Zuccotti${}^{5}$}

\affiliation{
\mbox{${}^1$ INAF, Osservatorio Astrofisico di Catania, via S.Sofia 78, I-9    5123 Catania, Italy  }\\
\mbox{${}^2$ INFN, Sezione di Catania,  via S. Sofia 64, I-95123,Catania, Italy.}\\
\mbox{${}^3$ Universit\`a degli Studi di Trento,\\Via Sommarive, 14, IT-38123, Trento, Italy} \\
\mbox{${}^4$ INFN - TIFPA,\\Via Sommarive, 14, IT-38123, Trento, Italy} \\
\mbox{${}^5$ Universit\`a di Pisa,\\ Largo Bruno Pontecorvo, 3, IT-56127 Pisa, Italy}
}

\begin{abstract}
The inclusion of the Weyl squared term in the gravitational action is one of the most simple, yet non trivial modifications to General Relativity at high energies. Nevertheless the study of the spherically-symmetric vacuum solutions of this theory
has received much attention only in recent times. A new type of asymptotically flat 
wormhole which does not match symmetrically 
at a finite radius with another sheet of the spacetime is presented. 
The outer spacetime is characterized by a newtonian potential with a Yukawa correction, and has gravitational properties that can be arbitrarily close to the ones of a Schwarzschild black hole. The internal spacetime instead possesses a singularity at $r=\infty$ with the topology of a 2-dimensional
sphere. The expansion scalar of geodesics reaching this singularity diverges in a finite amount of proper time, with a striking resemblance with the future singularity of the Big Rip cosmological scenario. In terms of the external Yukawa hair and mass $M$,
these new wormholes fill a large region of the two-dimensional parameter space of physical solutions with $M>0$. On the contrary black holes, both of Schwarzschild and non-Schwarzschild nature, are confined on a line. We argue that this type of wormholes are ideal black hole mimickers.
\end{abstract}

\maketitle

\section{Introduction}

The recent discovery of a new class of black holes \cite{Lu:2015cqa} has shown 
that the spectrum of classical solutions derived from  quadratic corrections 
to the classical Einstein-Hilbert lagrangian   is far from being clearly understood. 
The case of Einstein-Weyl theory, defined by 
\begin{equation}\label{sew}
S = \int{ d^4 x \sqrt{-g} [\gamma R -\alpha C_{\mu \nu \rho \sigma}C^{\mu \nu \rho \sigma}]}
\end{equation}
is particularly instructive. Spherically symmetric,  asymptotically flat solutions with vanishing Ricci scalar $R$ and non-zero Ricci
tensor $R_{\mu\nu}$ posses regular horizons and obey the first law of thermodynamics \cite{Lu:2015cqa}.
However, in spite of much analytical  \cite{Podolsky:2018pfe,Podolsky:2019gro,Saueressig:2021wam} and numerical \cite{Lu:2015psa,Goldstein:2017rxn} efforts
the role played by the asymptotic field in determining 
the global property of the solutions and the structure of the singularity 
still deserves further investigation \cite{Bonanno:2019rsq}.

This scenario has to be contrasted with the more familiar one in $f(R)$ theories, 
as in this case spherically symmetric solutions have been extensively studied in the past 
\cite{2011PhR...509..167C,Olmo:2019flu}, along with their stability properties in 
various astrophysical contexts \cite{Capozziello:2015yza,Astashenok:2013vza}.

The inclusion of the squared Weyl tensor term to the standard Einstein-Hilbert lagrangian
has a long history. Albeit originally motivated by the possibility  of perturbative
renormalization \cite{Stelle:1976gc}, in more recent times its  presence has
emerged \cite{machado,Benedetti:2009rx,Hamada:2017rvn} in the framework of the Asymptotically Safe program for Quantum Gravity \cite{Percacci:2017fkn,Reuter:2019byg} and of the
fakeons theory \cite{anselmi17,Anselmi:2018ibi}.
In this work we shall not discuss the viability of the theory defined in (\ref{sew})  at the 
quantum level because we would like to focus on the classical content of the theory extending 
the investigation started in \cite{Bonanno:2019rsq} to include 
wormhole solutions.
In fact, although a wormhole class of solutions have been found in \cite{Lu:2015psa}, their weight in the space of all the
possible spherically symmetric and asymptotically flat solutions, and their global properties are still unknown.
Indeed the so called ``non-symmetric" wormholes (no-sy WHs) found
in \cite{Lu:2015psa} cannot be analytically described in simple terms and their interpretation is still elusive. 
In this work we shall fill this gap and study this class of solutions in detail. 

The numerical approach described
in \cite{Bonanno:2019rsq} for black holes is clearly advantageous also in the case of wormholes, allowing to investigate the dependence of the local metric coefficient 
on the asymptotic field, which is described by the sum of a Schwarzschild and Yukawa term, in a systematic manner. In particular, it is possible to
determine the properties of the metric close to the wormhole throat as a function of the mass and the Yukawa coefficient at large distances. Once the metric functions and their derivatives have been determined at the throat radius, it is possible to extend the integration to the other patch of spacetime and characterize its behaviour at large distances. We show that these wormholes connect an asymptotically flat metric with a singular spacetime where the temporal and radial components of the metric vanish exponentially at infinity. 
This new type of singularity is perceived by infalling observers in a similar way as the future singularity of the Big Rip cosmological scenario \citep{Caldwell:1999ew}, with the expansion scalar of congruences of geodesics going to an infinite positive value in a finite amount of proper time. For external observers, instead, the singularity lies at the edges of the causal structure, and therefore is naked only in the infinite past limit. Having no trapped surfaces, however, these singularities are always avoidable with a sufficient amount of energy.
It is interesting to note that a similar spacetime has been found in \cite{Arrechea:2019jgx} as solution of General Relativity coupled with a quantum corrected stress-energy tensor, suggesting that no-sy WHs are indeed a common prediction of semiclassical theories of gravity.

At last, as the parameter space of possible solutions is getting understood \cite{Silveravalle:2022lid}, we also show that no-sy WHs densely populate this two-dimensional space, at variance with the BH class that is confined on two lines. In addition, no-sy WHs populate a large portion of the part of the parameter space with physical solutions with a positive mass, and are present for arbitrarily large masses. Schwarzschild black holes appear therefore as a limiting case of no-sy WHs in this large mass limit.

\section{Equations of motion and analytical approximations}
\subsection{Non-linear equations of motion}
The equations of motion of the Einstein-Weyl theory can be derived from the minimization of the action (\ref{sew}). In tensorial form they are written as
\begin{equation}\label{eomt}
\begin{split}
\mathcal{H}_{\mu\nu}&=\gamma\left(R_{\mu\nu}-\frac{1}{2}R\,g_{\mu\nu}\right)-4\alpha\left(\nabla^\rho\nabla^\sigma+\frac{1}{2}R^{\rho\sigma}\right)C_{\mu\rho\nu\sigma}\\
&=0,
\end{split}
\end{equation}
and, thanks to the traceless nature of the Weyl tensor, their trace is simply
\begin{equation}\label{trace}
\tensor{\mathcal{H}}{^\mu_\mu}\propto R=0.
\end{equation}
We now focus on static and spherically symmetric spacetimes, and choose the ansatz for the metric in Schwarzschild coordinates
\begin{equation}\label{ansmet}
ds^2=-h(r)\,dt^2+\frac{dr^2}{f(r)}+r^2d\Omega^2.
\end{equation}
Exploiting the trace equation (\ref{trace}) we can recast the equations of motion (\ref{eomt}) as a system of two second order ordinary differential equations in $h(r)$ and $f(r)$, as already shown in \citep{Lu:2015psa,Bonanno:2019rsq}. Explicitly these equations are
\begin{equation}\label{eom}
\begin{split}
&4 h(r)^2 \big(r f'(r)+f(r)-1\big)-r^2 f(r) h'(r)^2\\
&+r h(r) \big(r f'(r) h'(r)
+2 f(r) \big(r h''(r)+2 h'(r)\big)\big)=0,\\[0.3cm]
&\alpha  r^2 f(r) h(r) \big(r f'(r)+3 f(r)\big) h'(r)^2\\
&+2 r^2 f(r) h(r)^2 h'(r) \big(\alpha  r f''(r)+\alpha  f'(r)-\gamma  r\big)\\
&+h(r)^3 \big(r \big(3 \alpha  r f'(r)^2-4 \alpha  f'(r)+2
   \gamma  r\big)\\
&-2 f(r) \big(4 \alpha +2 \alpha  r^2 f''(r)-2 \alpha  r f'(r)+\gamma  r^2\big)\\
&+8 \alpha  f(r)^2\big)-\alpha  r^3 f(r)^2 h'(r)^3=0.
\end{split}
\end{equation}
Even if the symmetries of the spacetime greatly simplify the equations, it is clear that the full non-linear system cannot be solved exactly, and either approximations or numerical methods have to be used. In this section we will show the analytical approximations needed to study numerically the full system of equations (\ref{eom}), and to extract their physical properties.

\subsection{Linearized equations and solutions in the weak field limit}

In this work we are interested in asymptotically flat spacetimes, \textit{i.e.} solutions that describe isolated objects without a cosmological constant. We can therefore consider the weak field limit at large distances, in which the metric is a perturbation of the Minkowski spacetime. As described in \citep{Stelle:1977ry,Lu:2015psa,Bonanno:2019rsq} we write the functions $h(r)$ and $f(r)$ as
\begin{equation}
h(r)=1+\epsilon\,V(r),\qquad f(r)=1+\epsilon\,W(r),
\end{equation}
and expand (\ref{eomt}) at linear order in $\epsilon$. It is convenient to consider the 
combinations
\begin{equation}\label{eoml}
\begin{split}
\tensor{\mathcal{H}}{^\mu_\mu}&\propto\nabla^2 V(r)+2Y(r)=0,\\
\tensor{\mathcal{H}}{^i_i}-\tensor{\mathcal{H}}{^t_t}&\propto\left(\nabla^2-\frac{3\gamma}{4\alpha}\right)\nabla^2V(r)-\nabla^2Y(r)=0,
\end{split}
\end{equation}
where $Y(r)=r^{-2}\left(rW(r)\right)'$, which can be easily solved using Fourier modes. Imposing asymptotic flatness and fixing a parameterization of time (\textit{i.e.} imposing $h(r)\to 1$ as $r\to+\infty$) we can suppress some of the free parameters of the solution of (\ref{eoml}) and obtain
\begin{equation}\begin{split} \label{soluzioni linearizzate}
h(r)=\, &1-\frac{2\,M}{r}+2S^-_2 \frac{e^{-m_2\, r}}{r}, \\
f(r)=\, &1-\frac{2\,M}{r}+S^-_2 \frac{e^{-m_2\, r}}{r}(1+m_2\, r) ,
\end{split}\end{equation}
with $m_2^2=\frac{\gamma}{2\alpha}$ and $M$ being the ADM mass in Planck units, which is the Schwarzschild solution with exponentially suppressed corrections. We note that in the Newtonian limit the gravitational potential $\phi(r)\sim\frac{1}{2}(h(r)-1)$ will have a Yukawa correction, as expected for a massive mediator of the interaction, which can be either attractive or repulsive according to the sign of the ``charge'' $S_2^-$. We will see in section \ref{sec3} that the sign and the relative values of $M$ and $S_2^-$ will be crucial for having a no-sy WH type of solution.

\subsection{Series expansion at finite radii}

\begin{table*}[t]
\centering
\caption{Families of solutions around finite and zero radii in Einstein-Weyl gravity}
\begin{tabular}{c c c}\hline\hline\\[-0.27cm]
\phantom{aaaaaa}Family\phantom{aaaaaa} & \phantom{aaaaaa}$\mathrm{N}^{\mathrm{o}}$ of free parameters\phantom{aaaaaa} & \phantom{aaaaaaa}Interpretation\phantom{aaaaaaa}\\[0.05cm] \hline\\[-0.25cm]
$(0,0)_0^1$  & $2\,(\to 0)$ & Regular solution/True vacuum\\ 
$(-1,-1)_0^1$ & $3\,(\to 1)$ & Naked singularity/Schwarzschild interior\\ 
$(-2,2)_0^1$ & $4\,(\to 2)$ & Bachian singularity/Holdom star\\ 
$(0,0)_{r_0}^1$ & $4\,(\to 2)$ & Regular metric\\
$(1,1)_{r_0}^1$ & $3\,(\to 1)$ & Black hole\\
$(1,0)_{r_0}^1$ & $2\,(\to 0)$ & Symmetric wormhole\\
$(1,0)_{r_0}^2$  & $4\,(\to 2)$ & Non-symmetric wormhole\\
$(4/3,0)_{r_0}^3$ & $3\,(\to 1)$ & Not known\\[0.1cm]
\hline \hline
\end{tabular}
\label{tabfa}
\end{table*}
At finite radii we can expect the solutions to be well approximated by series expansions.
The different families of solutions allowed by the equations of motion have been exhaustively studied and classified in \citep{Lu:2015psa,Podolsky:2019gro} using a variation of the Frobenius method. Taking an expansion of the metric functions in the form
\begin{equation}
\begin{split}
h(r)&=\left(r-r_0\right)^t\!\left[\displaystyle\sum_{n=0}^N h_{t+\frac{n}{\Delta}}\left(r-r_0\right)^{\frac{n}{\Delta}}\!+O\!\left(\left(r-r_0\right)^{\frac{N\!+\!1}{\Delta}}\right)\!\right]\!,\\
f(r)&=\left(r-r_0\right)^s\!\left[\displaystyle\sum_{n=0}^N f_{s+\frac{n}{\Delta}}\left(r-r_0\right)^{\frac{n}{\Delta}}\!+\!O\left(\left(r-r_0\right)^{\frac{N\!+\!1}{\Delta}}\right)\!\right]\!,
\end{split}
\end{equation}
it is possible to classify the solutions as $(s,t)_{r_0}^\Delta$. The families known at present time are shown in Table \ref{tabfa}, where in the second column we made manifest the number of free parameters after imposing asymptotic flatness and a specific time parameterization. We note that in \citep{perkinsthesis} a non-Frobenius family with logarithmic corrections to the $(-1,-1)_0^1$, and with one additional free parameter, has been found and there are hints that it might populate a large area of the parameter space. We also specify that there are some differences in the notation used in our work and in \citep{Lu:2015psa,Podolsky:2019gro}, the main being the different sign for the exponent $s$ due to their metric ansatz in terms of the function $A(r)=1/f(r)$ for the families around $r_0=0$.\\[0.2cm]
As specified in the introduction, our main goal is to describe the physical properties of the non-symmetric wormhole type of solutions, that is solutions belonging to the $(1,0)_{r_0}^2$ family; having the maximum number of free parameters allowed in the theory, this family is expected to populate a non-zero measure region of the parameter space. We highlight here the behaviour of the metric around the ``throat'' $r=r_\wt$
\begin{equation}\begin{split}\label{wh einstein weyl}
h(r)=\, &h_0+h_{1/2}(r-r_\wt)^{\frac{1}{2}}+h_1 (r-r_\wt)+O((r-r_\wt)^{\frac{3}{2}}) ,\\[0.2cm]
f(r)=\, &f_1 (r-r_\wt)+ f_{3/2} (r-r_\wt)^{\frac{3}{2}}+O((r-r_\wt)^2),
\end{split}\end{equation} 
which is characterized by a divergent radial component of the metric $g_{rr}=1/f(r)$ and a regular, but with divergent derivative, temporal component of the metric $g_{tt}=-h(r)$. Moreover, this family of solution is completely determined by the four parameters $(r_\wt,h_0,h_{1/2},f_1)$, of which only specific combinations will lead to asymptotically flat solutions.

\subsubsection{Wormhole behaviour of $(1,0)_{r_0}^2$ solutions}

The metric in (\ref{wh einstein weyl}) is defined only for radii $r>r_\wt$, but having regular curvature invariants at $r=r_\wt$ we expect this to be a coordinates artefact. The simple extension to the region $r<r_\wt$ with an expansion in terms of semi-integer powers of $(r_\wt-r)$, suffers from a severe pathology, namely we have to choose between having discontinuities in the curvature invariants or having two temporal coordinates. The other possible extension is to consider a wormhole type of spacetime, where we join two $r>r_\wt$ patches at the ``throat'' $r=r_\wt$. We believe that this is the most sensible choice.\\
The wormhole nature of these type of solution is manifest after the coordinate transformation
\begin{equation}\label{rho transformation}
r=r_\wt+\frac{1}{4}\rho^2,
\end{equation}
in which the metric has the form
\begin{equation}\label{whrho}
\begin{split}
ds^2=\, & -h_0\left(1+\frac{h_{1/2}}{2}\rho+O\left(\rho^2\right)\right)dt^2\\
&+\frac{d\rho^2}{f_1+\frac{f_{3/2}}{2}\rho+O\left(\rho^2\right)}+\left(r_\wt+\frac{1}{4}\rho^2\right)^2d\Omega^2.
\end{split}
\end{equation}
This is manifestly well behaved around $\rho=0$, and then we can extend the metric to the $\rho<0$ region, corresponding to a second $r>r_\wt$ patch of a wormhole-type spacetime \citep{Lu:2015psa}. In contrast with standard wormhole solutions, however, the metric (\ref{whrho}) is not symmetric under the parity operation $\rho\to -\rho$.
Going back to the $r$ coordinate, we see that the metric on other side of the throat can be described switching the sign of the semi-integer terms in (\ref{wh einstein weyl}), that is
\begin{equation}\begin{split}\label{whewos}
h(r)=\, &h_0-h_{1/2}(r-r_\wt)^{\frac{1}{2}}+h_1 (r-r_\wt)-O((r-r_\wt)^{\frac{3}{2}}) ,\\[0.2cm]
f(r)=\, &f_1 (r-r_\wt)- f_{3/2} (r-r_\wt)^{\frac{3}{2}}+O((r-r_\wt)^2).
\end{split}\end{equation}
Looking at the explicit form of the parameters it can be proved that the expansion (\ref{whewos}) is exactly the one in (\ref{wh einstein weyl}) after the substitution $h_{1/2}\to -h_{1/2}$.\\

\subsection{Solution around vanishing metric}

Given the wormhole nature of the metric in (\ref{wh einstein weyl}), we present a spacetime characterized by two regions both mapped by $r\in [r_\wt,+\infty )$. It is natural to ask how the metric behaves in the second patch, and in particular if it can be asymptotically flat in both the spacetime patches. \\
From our numerical work we found a common behavior:
once imposed asymptotic flatness in the first patch, the metric in the second patch results to be non-asymptotically flat, with $f(r)$ diverging and $h(r)$ vanishing for $r\to+\infty$, corresponding to both $g_{rr}$ and $g_{tt }$ vanishing at large radius. At the present time, non-asymptotically flat solutions are expected in Einstein-Weyl gravity, but there is no analytical approximation of such solutions.\\
Following what we have from the numerical results we proceed by looking for a metric with $g_{rr}(r)$ and $g_{tt}(r)$ vanishing for $r\to +\infty$. In order to do this we expand the metric functions as
\begin{equation}\label{espansione asintotica}
\begin{split}
h(r)=\, &\epsilon\,h_1(r)\big(1+\epsilon\,h_2(r)+O(\epsilon^2)\big),\\ 
f(r)=\, &\frac{1}{\epsilon\,f_1(r)\big(1+\epsilon\,f_2(r)+O(\epsilon^2)\big)},
\end{split}
\end{equation}
then we can solve our equations of motion (\ref{eom}) order by order in $\epsilon$, starting from $O\left(\epsilon^{-1}\right)$.
An exact solution of the equations at first order can be found, and it has the form
\begin{equation}\label{comportamentoesponenziale}
\begin{split}
h_1(r)&=C_h\, \mathrm{e}^{-a\,r} r^2\\
f_1(r)&=C_f\, \mathrm{e}^{-a\,r} r^2
\end{split}
\end{equation}
in which $a,\, C_h,\, C_f$ are free parameters, with the constraint $a>0$, in order to be consistent with the initial assumption $g_{tt},g_{rr}\to0$ for $r\to+\infty$.
Now it is convenient to rewrite the ansatz (\ref{espansione asintotica}) as 
\begin{equation}
\begin{split}
\label{soluzione asintotica serie}
h(r)&=C_h\, \mathrm{e}^{-a\,r} r^2(1+\tilde{h}_2(r)\mathrm{e}^{-a\,r}+O(\mathrm{e}^{-2 a\,r}))\\
f(r)&=\frac{1}{C_f\, \mathrm{e}^{-a\,r} r^2(1+\tilde{f}_2(r)\mathrm{e}^{-a\,r}+O(\mathrm{e}^{-2 a\,r}))},
\end{split}
\end{equation}
where the expansion in $\epsilon$ is substituted by an expansion in the variable $y=\mathrm{e}^{-a\,r}$. When expanding (\ref{eom}) order by order in $y$, the first orders become a system of second order linear differential equations in $\tilde{h}_2(r)$ and $\tilde{f}_2(r)$. The solution can be found in a polynomial form 
\begin{equation}
\begin{split}
\label{secondo ordine soluzione asintotica }
\tilde{h}_2(r)&=\tilde{h}_0+\tilde{h}_1 \,r +\tilde{h}_2\, r^2+\tilde{h}_3\, r^3,\\
\tilde{f}_2(r)&=\tilde{f}_0+\tilde{f}_1\, r +\tilde{f}_2 \,r^2+\tilde{f}_3\, r^3,
\end{split}
\end{equation}
where $\tilde{f}_1$ result to be another free parameter and the other coefficients are completely determined by the four free parameters $(C_f,\,C_h,\,a,\,\tilde{f}_1)$.
Similarly the functions $\tilde{h}_3(r)$ and $\tilde{f}_3(r)$ at order $O\left(\mathrm{e}^{-3 a r}\right)$ can be found, obtaining two sixth degree polynomials in $r$ but with no other free parameters appearing. The total number of free parameters of these solutions results then to be four, the correct number needed to connect these solutions with the $(1,0)_{r_0}^2$ family.  \\
Although we do not discuss the convergence of this expansion, we believe that a certain convergence radius $r^*$ exists, such that for $r>>r^*$, the solution is well approximated by  (\ref{soluzione asintotica serie}).
The numerical behavior found with our data is in agreement with this expansion already at first order, as we will show in subsection \ref{sec4}.

\section{Numerical results and the structure of the spacetime}

\subsection{Metric characterization with the shooting method}

The shooting method has proven to be extremely useful for extracting the relevant physical properties of numerical solutions in quadratic gravity \citep{Bonanno:2019rsq,Bonanno:2021zoy}. We consider the metric to be described by the weak field limit approximation (\ref{soluzioni linearizzate}) at large distances ($r=r_{\infty}$), and by the series expansion (\ref{wh einstein weyl}) at sixth order close to the throat ($r=r_\wt+r_\epsilon$). The equations (\ref{eom}) are then numerically integrated with guessed values of the parameters $(M,S_2^-,r_\wt,h_0,h_{1/2},f_1)$ from both boundaries to a fitting radius $r=r_f$, where the continuity of $h(r)$, $f(r)$ and their derivatives is imposed with the use of a root finding algorithm. Once the convergence is achieved, this procedure fixes the value of the parameters $(r_\wt,h_0,h_{1/2},f_1)$ that set the initial condition (\ref{whewos}) at the other side of the throat for an additional integration towards $r\to +\infty$ (but with inverted sign for $\rho$), allowing us to study the second patch of the spacetime.\\
Throughout this analysis we integrated the equations using an adaptive stepsize method which switches between a midpoint and an implicit Euler methods, according to the stiffness of the system, with a tolerance of $10^{-14}$, implemented using the \texttt{Wolfram} language. The continuity of the metric is obtained through the Newton's method used by the \texttt{FindRoot} function defined in this language, with a precision of $10^{-4}$. The large distance radius has been fixed as $r_\infty=15$ in order to have Yukawa corrections greater than the tolerance threshold, and the distance from the throat as $r_\epsilon=10^{-3}$ in order to discard terms smaller than such threshold. The precise value of the fitting radius does not affect the accuracy of the shooting method, but is crucial for obtaining convergence efficiently; a value sufficiently close to the throat, in particular we used $r_f=r_\wt+5\cdot 10^{-2}$, has been found optimal for our purposes. We specify here that the equations have been rescaled in terms of the Spin-2 particle mass $m_2$ in order to have dimensionless quantities, and then all the scales of the solutions will be determined by the only free parameter of the theory $\alpha$; the length unit will be $l_2=4\sqrt{2\pi\alpha}\,l_p$, and the mass unit will be $m_2=(4\sqrt{2\pi\alpha})^{-1}m_p$. For future reference, we remember that with the notation used in (\ref{soluzioni linearizzate}) the Schwarzschild mass parameter $M$ has the dimension of length.

\begin{figure*}[t]
\centering
\includegraphics[width=\textwidth]{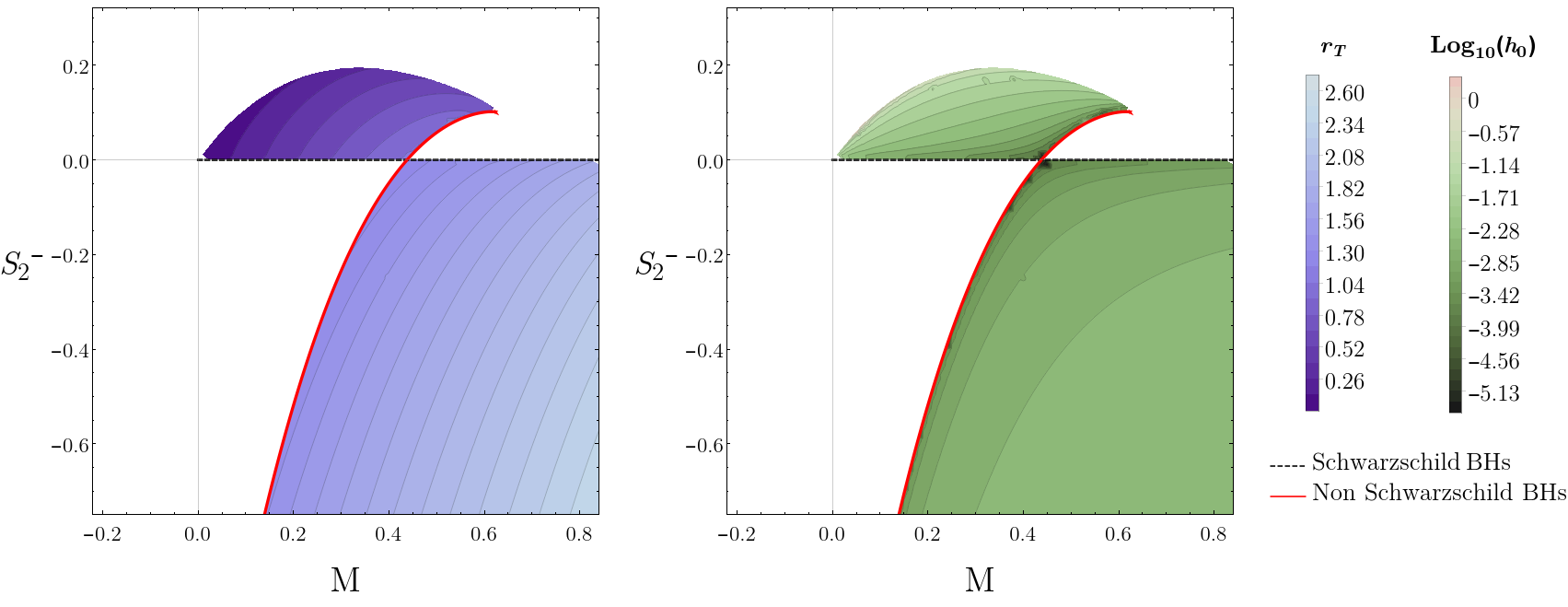}
\caption{Trend of the two main throat parameters in the phase diagram. In function of the gravitational parameters $M$ and $S_2^-$ we show the throat radius $r_\wt$ in the left panel, and the redshift parameter $h_0$ in the right panel.}
\label{whfam}
\end{figure*}

While for the one-parameter families described in \citep{Bonanno:2019rsq,Bonanno:2021zoy} it was sufficient to fix either the event horizon or the stellar surface radii to find convergence for all the other parameters, the two degrees of freedom of no-sy WHs require some preliminary steps in our procedure. First of all, we have optimal convergence by fixing at the beginning of the integration the weak field parameters $M$ and $S_2^-$ instead of some throat parameter (\textit{e.g.} the throat radius). We can mark the limits of the area in the $M$-$S_2^-$ parameter space where no-sy WHs can be found with an exploratory scan of this space. At practical level this means that the equations of motion are integrated using (\ref{soluzioni linearizzate}) as initial conditions with values of $M$ and $S_2^-$ chosen on a grid, and the values for which the equations are singular at a finite radius $r>0$ are saved in order to be used in the shooting method. As a cross check, the code where the shooting method is implemented never reaches convergence whenever initial values of $M$ and $S_2^-$ different from the one found with this scan are used. This preparatory integration is also used to find an educated guess for the parameters to use at the beginning of the shooting method, with a great increase in convergence efficiency. The preliminary scan of the parameter space is not only useful for improving convergence, but gave us relevant insight on the ``phase diagram'' of the theory, that is how the parameters $M$ and $S_2^-$ of the solutions determine the family they belong to. We will present the phase diagram of Einstein-Weyl gravity in further work.

\subsection{The family of no-sy WH solutions}\label{sec3}

Before discussing the details of no-sy WH solutions, we would like to present the global trend for some of the parameters of the $(1,0)_{r_0}^2$ family. In particular we believe that the most informative parameters are the gravitational ones, $M$ and $S_2^-$, the throat radius $r_\wt$, and the value of the temporal component of the metric at the throat $h_0$, that can be linked to the redshift of a photon emitted at such distance and measured at infinity by $z\left(r_\wt\right)=1/\sqrt{h_0}$. In figure \ref{whfam} we simultaneously show the area of the phase diagram populated by no-sy WHs, and the relations between the throat parameters $r_\wt$, $h_0$ and the gravitational ones $M$, $S_2^-$. No-sy WHs are found in two distinct regions: for a repulsive contribution of the Yukawa term in the potential (\ref{soluzioni linearizzate}), \textit{i.e.} $S_2^->0$, the region of no-sy WHs is delimited by the Schwarzschild and non Schwarzschild black holes lines, and has a smooth transition into a region populated by horizonless solutions belonging to a logarithmic correction of the $(-1,-1)_0^1$ family; for an attractive contribution of the Yukawa term, \textit{i.e.} $S_2^-<0$, the region is still delimited by the Schwarzschild and non Schwarzschild black holes lines, but is unbounded for large $M$ and $S_2^-$. In both cases black holes appear as a transition between no-sy WHs and solutions of the $(-2,2)_0^1$ family. In particular we would like to stress the fact that, if we consider only positive mass solutions as physical, \textit{no-sy WHs populate almost half of the physical region of the phase diagram}. If there are no criteria, \textit{e.g.} symmetry arguments, for selecting solutions with $S_2^-=0$, it is then natural to consider no-sy WHs as much more viable candidates than Schwarzschild black holes as generic vacuum solutions, even in the large mass limit.\\[0.15cm]
The relations of the throat radius $r_\wt$ and the redshift parameter $h_0$ are useful consistency checks. We see from figure \ref{whfam} that the throat radius increases with an increasing mass, and in particular is consistent with the Schwarzschild mass-radius relation $r_H=2\,M$ as the Yukawa charge goes to zero. We note that the throat radius increases also as the Yukawa charge decreases, and then in the large mass limit, where no-sy WHs are present only for negative values of $S_2^-$, no-sy WHs are always larger than the Schwarzschild black hole with the same mass. At last, as the gravitational parameters of a no-sy WH gets closer to the ones of a black hole, be it either of the Schwarzschild or non Schwarzschild families, the redshift of a photon emitted at the throat increases, and the topological sphere defined by $r_\wt$ becomes an infinite redshift surface in this limit. In other words, for large masses, if the Yukawa charge is sufficiently small no-sy WHs are optimal black hole mimickers.

\begin{figure*}[t]
\centering
\includegraphics[width=\textwidth]{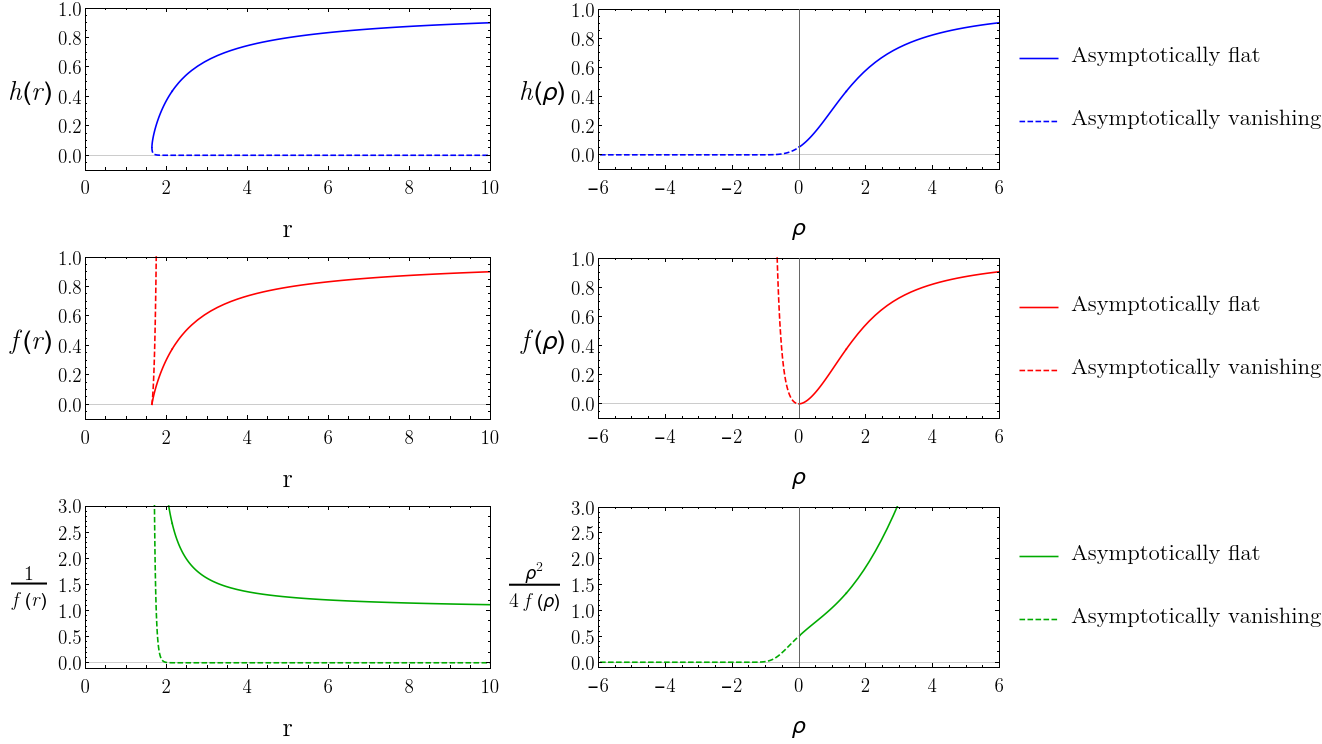}
\caption{Metric of a no-sy WH solution with $M=0.5$ and $S_2^-=-0.3$: in the panels on the left the metric is in function of the $r$-coordinate, while in the panels on the right is in function of the $\rho$-coordinate; solid and dashed lines indicate whether we are in the asymptotically flat or in the asymptotically vanishing patch, respectively.}
\label{whmetric}
\label{whcompletorho}
\label{whcompletor}
\end{figure*}

\subsection{The no-sy WH spacetime}\label{sec4}

\begin{figure}[h]
\centering
\includegraphics[width=\columnwidth]{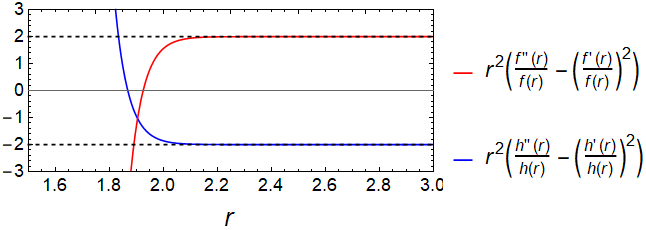}
\caption{Details of the metric in the asymptotically vanishing region of a no-sy WH solution with $M=0.5$ and $S_2^-=-0.3$.\label{plotandamentoasintotico}}
\end{figure}

We discuss here the main features of the no-sy WH solutions. The numerical results of the shooting method allowed us to characterize the no-sy WHs in the whole available space $\rho \in (-\infty,+\infty)$. We present the common behavior  found for all the solutions studied, except for a small area located in the region of positive Yukawa charge. These exceptions correspond to solutions with the wormhole throat less than a maximum radius $r_\wt\le\frac{1}{\sqrt{3}}$, in our unit, so we expect them to appear only at microscopical scales. For simplicity we do not discuss the behavior of these exceptions, instead we focus on all the other solutions, recalling that, on the contrary, they can have an arbitrary large radius.\\
In what follows, we have chosen the region of spacetime with positive $\rho$ as the asymptotically flat region. With this choice for $r\to+\infty$, $\rho\to+\infty$, the metric is determined by (\ref{soluzioni linearizzate}) with the line element
\begin{equation}\begin{split}\label{metrica linearizzata}
ds^2=&-\Bigl(1-\frac{2\,M}{r}+2S^-_2 \frac{e^{-m_2\, r}}{r}\Bigr)dt^2\\
&+\frac{1}{1-\frac{2\,M}{r}+S^-_2 \frac{e^{-m_2\, r}}{r}(1+m_2\, r)}dr^2+r^2 d\Omega^2,
\end{split}\end{equation}
for $r\to r_\wt$, $\rho\to 0^\pm$, it is determined by (\ref{wh einstein weyl}),
(\ref{whewos}) with the line element
\begin{equation}\begin{split}\label{metrica rwt}
ds^2=&-h_0\Bigl(1\pm h_{1/2}(r-r_\wt)^{\frac{1}{2}}+O(r-r_\wt)\Bigr)dt^2\\
&+\frac{1}{f_1 (r-r_\wt)\pm O((r-r_\wt)^{\frac{3}{2}})}dr^2+r^2 d\Omega^2,
\end{split}\end{equation}
and for $r\to+\infty$, $\rho\to-\infty$, it is determined by (\ref{soluzione asintotica serie}) with the line element
\begin{equation}\begin{split}\label{metrica asintotica}
ds^2=&-C_h\, r^2 \mathrm{e}^{-a\,r}(1+O(\mathrm{e}^{- a\,r}))dt^2\\
&+C_f\, r^2\mathrm{e}^{-a\,r}(1+O(\mathrm{e}^{-a\,r}))dr^2+r^2 d\Omega^2.
\end{split}\end{equation}
In figure \ref{whmetric} an example of no-sy WH spacetime is presented.
In terms of the $\rho$-coordinate, the functions $h(\rho)$ and $f(\rho)$, as well as $g_{\rho \rho}(\rho)$ are smoothly matched from both the patches in $\rho=0$.\\
The function $h(\rho)$ results to be monotonic, meaning that an observer would feel a gravitational force always in direction of decreasing $\rho$. This corresponds to an attractive central force in the asymptotically flat patch and a repulsive central force in the second patch.\\
\begin{figure}
\centering
\includegraphics[width=\columnwidth]{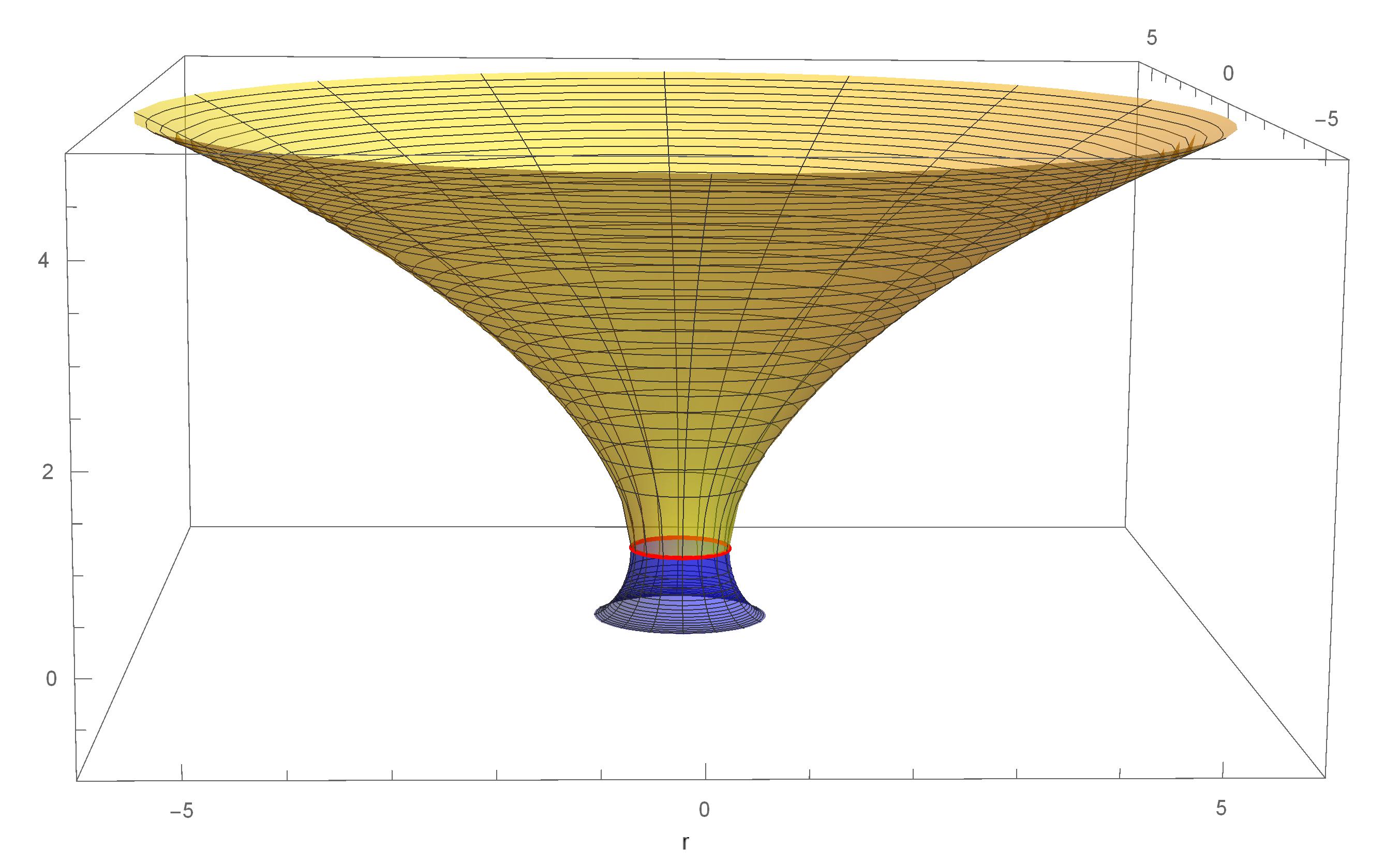}
\caption{Embedding diagram of a no-sy WH solution with $M=0.55$ and $S_2^-=0.14$: the yellow part corresponds to the $\rho>0$ region, the red curve corresponds to $r=r_\wt$, the blue part corresponds to the $\rho<0$ region.}
\label{embedding}
\end{figure}
In figure \ref{embedding} we show the embedding diagram of no-sy WHs around $r=r_\wt$. 
This is built in order to have a radial displacement $dr$ on the horizontal plane corresponding to a displacement of  proper distance $d\tilde{r}=\frac{dr}{\sqrt{f(r)}}$ on the surface embedded. The embedding diagram is clearly similar to what we have for solutions with horizon, since in both cases $f(r)$ vanishes at a certain radius, however here the non symmetric behavior of such wormholes is explicit. Since $f(r)$ rapidly grows in the second patch, when $f(r)>1$ the proper distance $d\tilde{r}$ becomes shorter than $dr$ and the spacetime cannot be further embedded. \\
Finally we show the details of the metric for $\rho<0$. In the previous plots the metric functions seem to have an exponential character in this region. We considered the ratio between the metric functions and their first derivatives. In particular in figure \ref{plotandamentoasintotico} we plotted the derivative of this ratio, which for an exact exponential is expected to vanish. We found the following limits 
\begin{equation}\begin{split}
\frac{d\ }{dr}\bigg(\frac{f'(r)}{f(r)}\bigg)&=\frac{f''(r)}{f(r)}-\bigg(\frac{f'(r)}{f(r)}\bigg)^2\to\frac{2}{r^2},\\
\frac{d\ }{dr}\bigg(\frac{h'(r)}{h(r)}\bigg)&=\frac{h''(r)}{h(r)}-\bigg(\frac{h'(r)}{h(r)}\bigg)^2\to-\frac{2}{r^2},
\end{split}\end{equation}  
that do not depend on the particular solution considered. 
This asymptotic behavior can be analytically integrated, finding again the first order of the asymptotically vanishing metric (\ref{soluzione asintotica serie}) and confirming that the solutions in the second patch are given by this expansion at large radius.\\  
The non-flat behavior (\ref{soluzione asintotica serie}) brings several implications for the spacetime structure of the second patch.\\
With such behavior the asymptotic surface $r\to +\infty$, $\rho\to -\infty$ results located at a finite proper distance from the wormhole throat. Indeed the proper radial distance is given by
 \begin{equation}
\tilde{r}_{max}=\int^\infty_{r_\wt}{\frac{dr }{\sqrt{f(r)}}},
\end{equation}
and this integral converges with $f(r)$ interpolated between (\ref{wh einstein weyl}) and (\ref{soluzione asintotica serie}).
The proper volume of the entire $\rho<0$ region is finite: it is given by
\begin{equation}
V_{p}=4 \pi \int^\infty_{r_\wt}{dr  \frac{r^2}{\sqrt{f(r)}}},
\end{equation}
that converges again.\\

From the monotonicity of $h(\rho)$, the surface $r\to+\infty$ also results to be attractive. \\
These considerations completely changes the nature of the second patch and highlights the non-symmetric nature of these solutions. In section \ref{section:singularity} we focus on the peculiar features of the metric at the surface $r\to+\infty$ of the second patch. 

\subsubsection{Geodesic dynamics and photon sphere}

Before we discuss the details of the asymptotic surface of the second patch, we show the geodesic dynamics of the no-sy WH spacetime. We start recalling that a general geodesic in a static spherically symmetric spacetime can be defined as the integral line of the vector field $V^\mu$ with components

\begin{equation}\label{geodetica generale}
\begin{split}
V^t=\frac{dt}{d\tau}&=\frac{E}{h(r)},\\
V^r=\frac{d r}{d \tau}&=\pm\sqrt{f(r)\Big(\frac{E^2}{h(r)}-\frac{L^2}{r^2}+\kappa\Big)},\\
V^\theta=\frac{d\theta}{d\tau}&=0,\\
V^\phi=\frac{d\phi}{d\tau}&=\frac{L}{r^2},
\end{split}
\end{equation}

with $\kappa=-1$ and $\tau$ the proper-time for a time-like geodesic, or $\kappa=0$ and $\tau$ an affine parameter for a null geodesic, and where we considered $\theta=\pi/2$ without loss of generality.\\
Firstly we note that the wormhole nature is manifest in the geodesic dynamics around the throat. Indeed, recalling the transformation (\ref{rho transformation}), the radial component in (\ref{geodetica generale}) can be written in terms of $\rho$ as

\begin{equation}\label{geodetica rho}
\begin{split}
\frac{d \rho}{d \tau}&=\pm \sqrt{\frac{f(r)}{(r-r_\wt)}\Big(\frac{E^2}{h(r)}-\frac{L^2}{r^2}+\kappa\Big)}.
\end{split}
\end{equation}
With $f(r)$ given by the metric around the throat (\ref{wh einstein weyl}), equation (\ref{geodetica rho}) can be integrated obtaining a smooth geodesic that goes from positive to negative $\rho$. Together with this, equation (\ref{geodetica generale}) tells that $\frac{dr}{d\tau}$ is forced to vanish at the throat. On the contrary, with this type of metric it is not possible to build a differentiable geodesic that goes from $r>r_\wt$ to $r<r_\wt$, confirming the interpretation of the $(1,0)_{r_0}^2$ as a wormhole solution family. \\
The second thing we want to highlight is that, if a free-falling object enters into the wormhole throat from the asymptotically flat patch it will proceed until it reaches the surface $r\to+\infty$ of the second patch.\\
In order to enter into the wormhole, the radical argument in (\ref{geodetica rho}) must be positive: around the throat we have $f(r)=f_1 (r-r_\wt)+O((r-r_\wt)^{\frac{3}{2}})$ so for positive $f_1$ we get the condition
\begin{equation}
\frac{E^2}{h(r_\wt)}>\frac{L^2}{r_\wt^2}-\kappa.
\end{equation}
Once entered in the second patch, $\frac{d\rho}{d\tau}$ cannot vanish, indeed we have
\begin{equation}
\frac{E^2}{h(r)}>\frac{E^2}{h(r_\wt)}>\frac{L^2}{r_\wt^2}-\kappa>\frac{L^2}{r^2}-\kappa,
\end{equation}
since $h(r)$ results decreasing in the asymptotically vanishing region. This means that a free-falling object will inevitably reach $\rho=-\infty$, since it is attracted by the gravitational force, and the angular momentum conservation contributes in the same direction. \\
Moreover this happens in a finite interval of proper time.
The proper time interval needed to fall into this surface is
\begin{equation}\label{protim}
\begin{split}
\tau_s=\int^{\infty}_{r(0)}{\frac{1}{ \sqrt{f(r)\Big(\frac{E^2}{h(r)}-\frac{L^2}{r^2}+\kappa\Big)}}dr},
\end{split}
\end{equation}
that is certainly convergent due to the asymptotic behaviour $\frac{f(r)}{h(r)}=O(\frac{e^{2 a r}}{r^4})$.\\
For a distant observer in the asymptotically flat patch instead, a particle falls into the wormhole throat in a finite time interval, but the time needed to reach the surface $\rho \to -\infty$ results divergent, as it is given by
\begin{equation}\label{tempo asintotico}
\begin{split}
t=\int^{r(t)}_{r(0)}\frac{E}{\sqrt{f(r)\Big(E^2\,h(r)-(\frac{L^2}{r^2}+\kappa)h(r)^2\Big)}}dr.
\end{split}
\end{equation}
Around the throat we have $h(r)=h_0+O(r-r_\wt)$ and $f(r)=f_1 (r-r_\wt)+O((r-r_\wt)^{\frac{3}{2}})$ so the integral in (\ref{tempo asintotico}) converges to a finite time interval for the distant observer.\\
On the other hand, with the asymptotic behavior (\ref{soluzione asintotica serie}) the integrand in (\ref{tempo asintotico}) tends to a constant but it must be integrated to infinity, so the integral diverges for $r(t)\to +\infty$.\\
The geodesic behavior around the throat implies that the photon sphere of no-sy WHs is always located in the asymptotically flat patch. For a distant observer in this patch, a no-sy WH appears like an ultra-compact object located inside its photon sphere. The asymptotic surface $\rho \to-\infty$ appears like an attractive horizon "inside" the throat. Instead any free-falling observer in the second patch falls into this surface in a finite amount of proper time but it can always spend energy to escape.

\section{The nature of the singularity and discussion \label{section:singularity}}

\begin{figure}[t]
\centering
\includegraphics[width=\columnwidth]{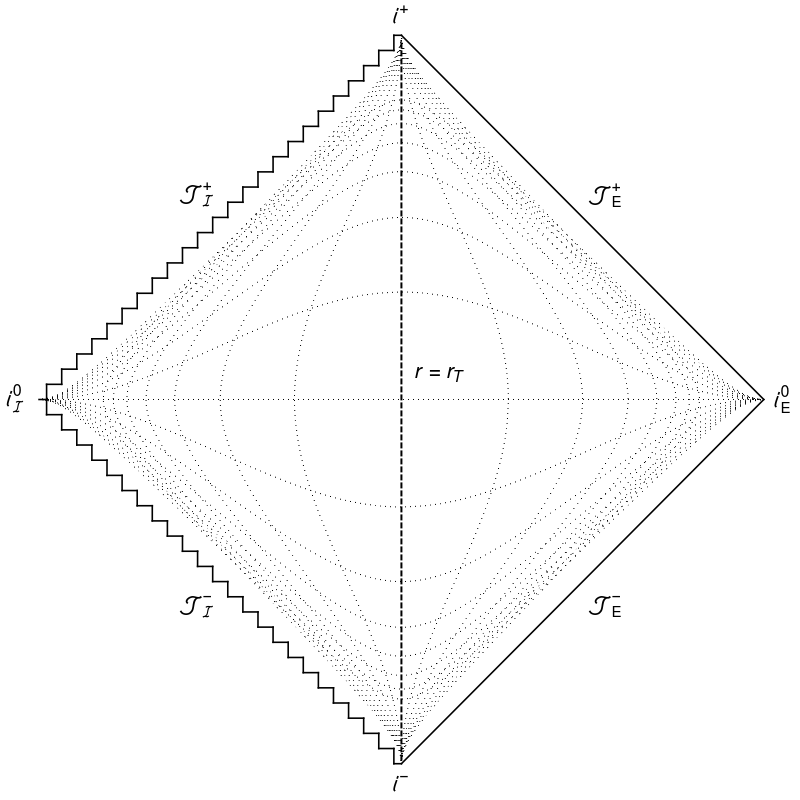}
\caption{Conformal diagram of a no-sy WH spacetime; the dotted lines indicate surfaces of constant time and radius.}
\label{penrose}
\end{figure}

The peculiar properties of the spacetime at the hypersurface defined by $r\to+\infty$ in the asymptotically vanishing patch strongly suggest that it should be a singular region. Indeed the behavior (\ref{soluzione asintotica serie}) implies the following limits for the curvature invariants: 
\begin{equation}\begin{split}
     R&=0 \qquad\qquad\ \, \mbox{for} \quad r\to+\infty,\\
     R_{\mu\nu}R^{\mu\nu}&=O\Big(\frac{e^{2ar}}{r^6}\Big) \quad \mbox{for} \quad r\to+\infty,\\
     R_{\mu\nu\rho\sigma}R^{\mu\nu\rho\sigma}&=O\Big(\frac{e^{2ar}}{r^6}\Big) \quad \mbox{for} \quad r\to+\infty;
\end{split}
\end{equation}
with the Ricci scalar being identically zero for our e.o.m., but with divergent squared Ricci tensor and Kretschmann scalar. Quite interestingly, the squared Weyl tensor results regular due to a cancellation between the divergent part of the Kretschmann scalar and the squared Ricci tensor. In particular using the precise result for $\tilde{h}_3(r)$ and $\tilde{f}_3(r)$ in (\ref{soluzione asintotica serie}) it is possible to prove the limit 
\begin{equation}
   C_{\mu \nu \rho \sigma}C^{\mu \nu \rho \sigma}=3m_2^4 \qquad\quad \mbox{for} \quad r\to+\infty,
\end{equation}
that does not depend on the particular solution considered. This result is crucial for the application of the finite action principle \citep{1988Natur.331...31B}, that is getting much attention in recent times for its applications in quadratic theories of gravity \citep{Lehners:2019ibe,Borissova:2020knn,Chojnacki:2021ves}. 
We also note that, beside the singular nature of the curvature invariants, the fact that a timelike geodesic reaches an infinite radius in a finite proper time is a strong indication that the spacetime is geodesically incomplete.\\
As it is clear that the hypersurface at $r\to+\infty$ is a singular region, it is a new kind of singularity with unique physical properties. To begin with, the causal structure of the spacetime is radically different from the standard solutions of General Relativity. As shown in the conformal diagram of figure \ref{penrose}, we see that the causal structure of a no-sy WH is equivalent to the one of a maximally extended Minkowski with a singularity at the ``internal'' $\mathcal{J}^+_I$ and $\mathcal{J}^-_I$. This is not surprising, considering that the $t$-$r$ sector of a no-sy WH spacetime is conformally equivalent to the Minkowski one after the coordinate transformation
\begin{equation}
ds^2=h(r)(-dt^2+dr^{*2})+r^2d\Omega^2,
\end{equation}
where the tortoise coordinate goes to zero at the throat. The relevant information that is manifestly shown in figure \ref{penrose}, however, is that the singularity is at the edges of the causal structure. In other words, a distant observer can communicate with the singularity only in an infinite amount of time. Furthermore, if we recall the definition of the redshift of a photon emitted at radius $r$ and measured at infinity $z(r)=\frac{1}{\sqrt{h(r)}}$, we see that the singularity is actually on an infinite redshift surface and, as for an event horizon, an infinite amount of energy is required to leave it. The singularity is therefore naked only in its infinite past section, and can be interpreted as the equivalent of a white hole singularity. The problem of dealing with naked singularities in no-sy WH spacetimes is then reduced to finding a collapse mechanism for the generation of such objects; however, we postpone this study to further work.\\
We now consider the behaviour of a congruence of infalling geodesics with tangent vector (\ref{geodetica generale}), where with infalling we mean that has a negative $V^r$ in the asymptotically flat patch, and a positive $V^r$ in the asymptotically vanishing one. Following the discussion in chapter 4 of \citep{Hawking:1973uf}, we consider the vector field $Z^\mu$, which represent the separation of points in nearby geodesics, that satisfies the equation
\begin{equation}\label{devequ}
\frac{\mathrm{d}}{\mathrm{d}\tau}Z^\mu=\tensor{B}{^\mu_\nu}Z^\nu,
\end{equation}
where we have defined the deviation tensor
\begin{equation}\label{devten}
B_{\mu\nu}=\tensor{h}{^\rho_\mu}\tensor{h}{^\sigma_\nu}\nabla_\rho V_\sigma
\end{equation}
with $h_{\mu\nu}$ being the metric of either the hypersurface orthogonal to the geodesic in the timelike case or the surface transverse to the geodesic in the null case. The first thing we want to highlight is that the expansion scalar
\begin{equation}
\theta=h^{\mu\nu}B_{\mu\nu}
\end{equation}
does not go to a \textit{negative} infinite value at the singularity, as it happens in the Schwarzschild case, but it goes to a \textit{positive} infinite value as in the limit of outgoing geodesics reaching spatial infinity in asymptotically flat solutions. However, the peculiar feature here is that \textit{the expansion scalar goes to infinity in a finite proper time}. It is in fact possible to prove that the expansion scalar at large radii satisfies
\begin{equation}
\theta(\tau)>\frac{1}{\left(\tau_s-\tau\right)^\alpha},
\end{equation}
with $0<\alpha<1$ and $\tau_s$ being the proper time (or the affine parameter) at which the geodesic reaches the singularity, and then that it diverges in the limit $\tau\to\tau_s$. The second thing we want to highlight is the behaviour of the deviation vector $Z^\mu$. If we restrict ourselves to radial geodesics, using the definition of proper time (\ref{protim}) and the asymptotic expansion (\ref{comportamentoesponenziale}) we can solve the differential equations (\ref{devequ}) close to the singular surface as
\begin{equation}\label{vecdev}
\begin{split}
Z^\mu=\begin{cases}
Z^t(r)\sim c^t\,r,\\
Z^r(r)\sim c^r\,r,\\
Z^\theta(r)\sim c^\theta\,r,\\
Z^\phi(r)\sim c^\phi\,r,
\end{cases}\qquad Z^\mu=\begin{cases}
Z^t(r)\sim c^t,\\
Z^r(r)\sim c^r,\\
Z^\theta(r)\sim c^\theta\,r,\\
Z^\phi(r)\sim c^\phi\,r,
\end{cases}
\end{split}
\end{equation}
in the timelike and null cases respectively. While null geodesics diverge only for geometrical aspects, timelike geodesics experience extreme tidal forces in the radial and temporal directions, that actually diverge as they get closer to the singularity. The presence of such disruption of timelike observers at a finite value of the proper time has a remarkable resemblance with the Big Rip cosmological scenario, where the expansion of the universe diverges in a finite amount of cosmological time. This Big Rip-like singularity is however localized inside a topological sphere of radius $r=r_\wt$ for an observer in the asymptotically flat patch, and has an ``origin'' in the topological sphere of radius $r=r_\wt$ for an observer in the asymptotically vanishing one.\\[0.2cm]
With the information at our disposal, we can now have an insight on how no-sy WHs are perceived by observers:
\begin{itemize}
\item[-] \textit{infalling} observers coming from the asymptotic flat patch are attracted by the no-sy WH just as by other compact objects, but after they have reached the radius $r=r_\wt$ they start to feel a repulsive force, and tidal forces in all directions: they are quickly pushed away to spatial infinity, and the tidal forces become so strong that are able to break all the binding energies and completely disrupt the observer in a finite amount of proper time, just like in the Big Rip cosmological scenario; however, in principle observers can always turn on a rocket and escape their fate;
\item[-] \textit{distant} observers in the asymptotically flat patch see an attractive object enclosed inside the topological sphere of radius $r=r_\wt$; the object is smaller than its photon sphere, and the light emitted outside this sphere will be absorbed by the object; however, particles can emit light from inside the object, but this emission is expected at extremely low frequencies: first of all photons are exponentially redshifted, and the temperature of a ball of gas is expected to decrease, as the volume increases as can be seen from (\ref{vecdev}); moreover, distant observers will never see the disruption of the infalling gas, that is instead perceived as ``frozen'' inside the object.
\end{itemize}
In conclusion, we can interpret no-sy WHs as black hole mimickers with a ``singularity by disruption'' istead of a ``singularity by compression'' that, for this reason, is always avoidable.

\section{Conclusions}

In this paper we give a complete description of the non-symmetric wormhole (no-sy WH) type of solutions of Einstein-Weyl gravity. With different analytical approximations, and using a shooting method procedure, we managed to link the properties of the spacetime at large distances to the ones close to the wormhole throat, and also to explore the second patch of the spacetime. No-sy WH solutions are characterized by an asymptotically flat patch, where they are described by a newtonian potential with a Yukawa correction, and by an asymptotically vanishing patch, where geodesics reach a singular surface in a finite proper time; the two patches are joined at the ``throat'', that is the topological sphere with the minimum radius that can be reached in this kind of spacetime. The values of the ADM mass and the Yukawa charge for which we have this type of solutions reveal that no-sy WHs populate a large area of the physical part of the parameter space of the theory, and that they are a much more viable candidate to be the generic vacuum solution of Einstein-Weyl gravity than black holes. The behaviour of the metric close to the throat, instead, suggests that no-sy WHs are optimal black hole mimickers, with the throat being an extremely high redshift surface, and with the photon sphere having a radius always larger than the throat. The singularity in the asymptotically vanishing patch has the peculiar behaviour of resembling the Big Rip cosmological singularity, with the presence of extreme tidal forces in all directions that completely disrupt timelike observers in a finite proper time. Despite having no horizons, the singularity results naked only in the infinite past, being at the edges of the causal structure of the solution. The absence of horizons, however, guarantees that there are no trapped surfaces, and with enough energy an outgoing geodesics can always escape from the singularity and reach the asymptotically flat region. All these properties suggest that no-sy WHs might be the substitutes predicted by quadratic gravity to black hole solutions. In order to make a definite statement, however, is fundamental to adress the stability of the solutions, that we plan to tackle in future, and subsequently the formation of no-sy WHs, their rotating counterparts, and the properties of an accretion disk surrounding them.

\acknowledgements S.S. would like to thank Massimiliano Rinaldi for his useful advice and stimulating discussions. This work has been partially supported by the INFN grant FLAG and the TIFPA - Trento Institute for Fundamental Physics and Applications.

\bibliographystyle{apsrev4-1}

\bibliography{whbib}

\end{document}